\documentclass[12pt,journal,compsoc]{IEEEtran}

\usepackage{graphicx}
\usepackage{float} 
\usepackage{subcaption}

\usepackage{footnote}
    
    \usepackage{printlen}

\ifCLASSOPTIONcompsoc
\else
\fi
\ifCLASSINFOpdf
\else
\fi
\begin{document}
%
\title{A Brief Comparison Between Available Bio-printing Methods}
%
%
%
%

\author{Ali~Bakhshinejad,Roshan M D'Souza

\IEEEcompsocitemizethanks{\IEEEcompsocthanksitem A. Bakhshinejad is a PhD student in Mechanical Engineering Department of University of Wisconsin-Milwaukee.\protect\\
E-mail: bakhshi3@uwm.edu
\IEEEcompsocthanksitem R.M. D'Souza is Associate Professor in mechanical engineering at university of Wisconsin-Milwaukee.}
\thanks{}}

\markboth{Great Lakes Biomedical Conference}%
{Bakhshinejad \MakeLowercase{\textit{et al.}}: A Brief Comparison Between Available Bio-printing Methods }
%



\IEEEcompsoctitleabstractindextext{%
\begin{abstract}
The scarcity of organs for transplant has led to large waiting lists of very sick patients. In drug development, the time required for human trials greatly increases the time to market. Drug companies are searching for alternative environments where the $in-vivo$ conditions can be closely replicated. Both these problems could be addressed by manufacturing artificial human tissue. Recently, researchers in tissue engineering have developed tissue generation methods based on 3-D printing to fabricate artificial human tissue. Broadly, these methods could be classified as laser-assisted and laser free. The former have very fine spatial resolutions (10s of $\mu$m) but suffer from slow speed ( $< 10^2$ drops per second). The later have lower spatial resolutions (100s of $\mu$ m) but are very fast (up to $5\times 10^3$ drops per second). In this paper we review state-of-the-art methods in each of these classes and provide a comparison based on reported resolution, printing speed, cell density and cell viability.  
\end{abstract}

}

\maketitle

\IEEEdisplaynotcompsoctitleabstractindextext

%
\IEEEpeerreviewmaketitle

\section{Introduction}

\IEEEPARstart{I}{ntroduction} of the first 3D printing method by Charles W. Hull in 1986 \cite{Hull1986}, changed the world of manufacturing. Complex shapes that previously could not be manufactured without expensive tooling such as mutli-part molds could be manufactured with ease through layered deposition of material. The same technique has now been adpated to manufacture complex human tissue. Over the last two decades, researchers have focused on techqniues to accomodate the sensitivity of live cells to stresses(friction, pressure, fluid viscosity etc.) that are ecountered during the printing process \cite{Guillemot2011,binder2011,ahn2014,Derby2008}.

\begin{figure*}
        \centering
        \begin{subfigure}[b]{0.3\textwidth}
                \includegraphics[width=\textwidth]{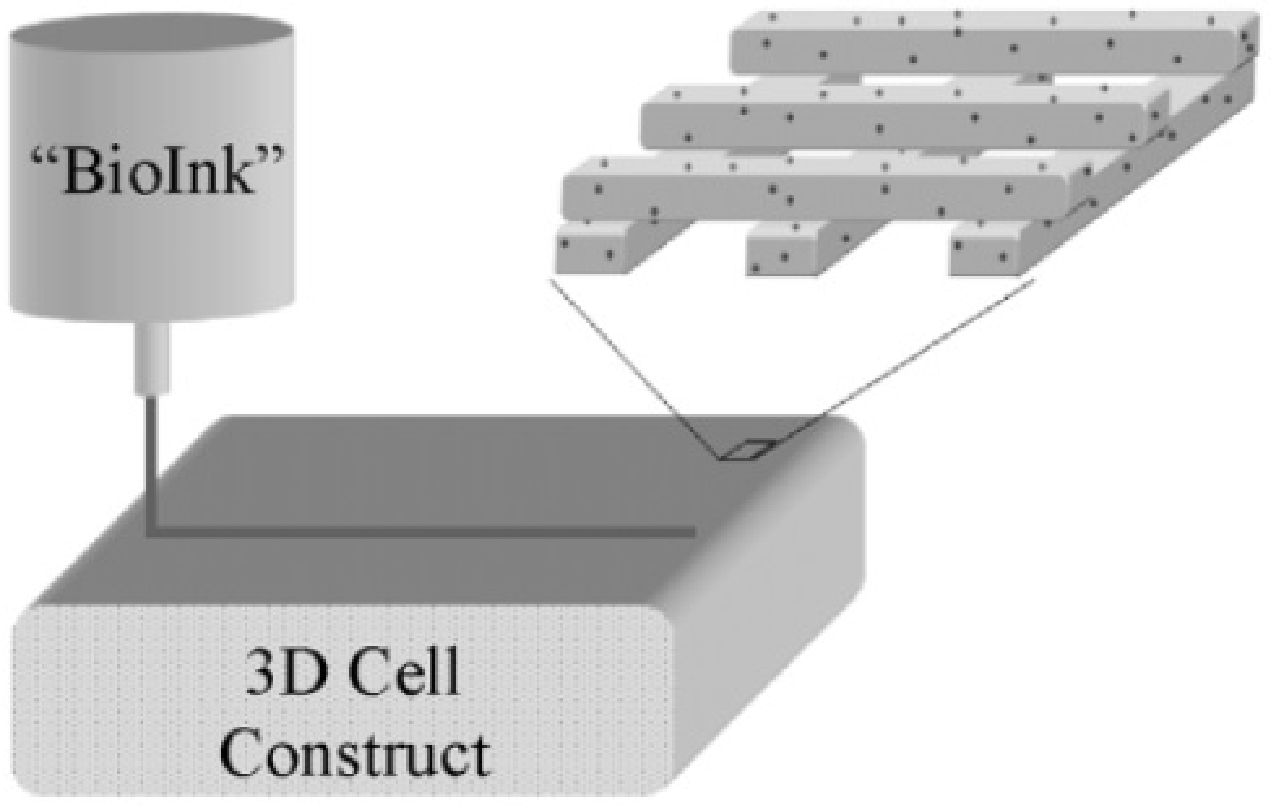}
                \caption{}
                 \label{fig:scaffold1}
        \end{subfigure}%
        \begin{subfigure}[b]{0.3\textwidth}
                \includegraphics[width=\textwidth]{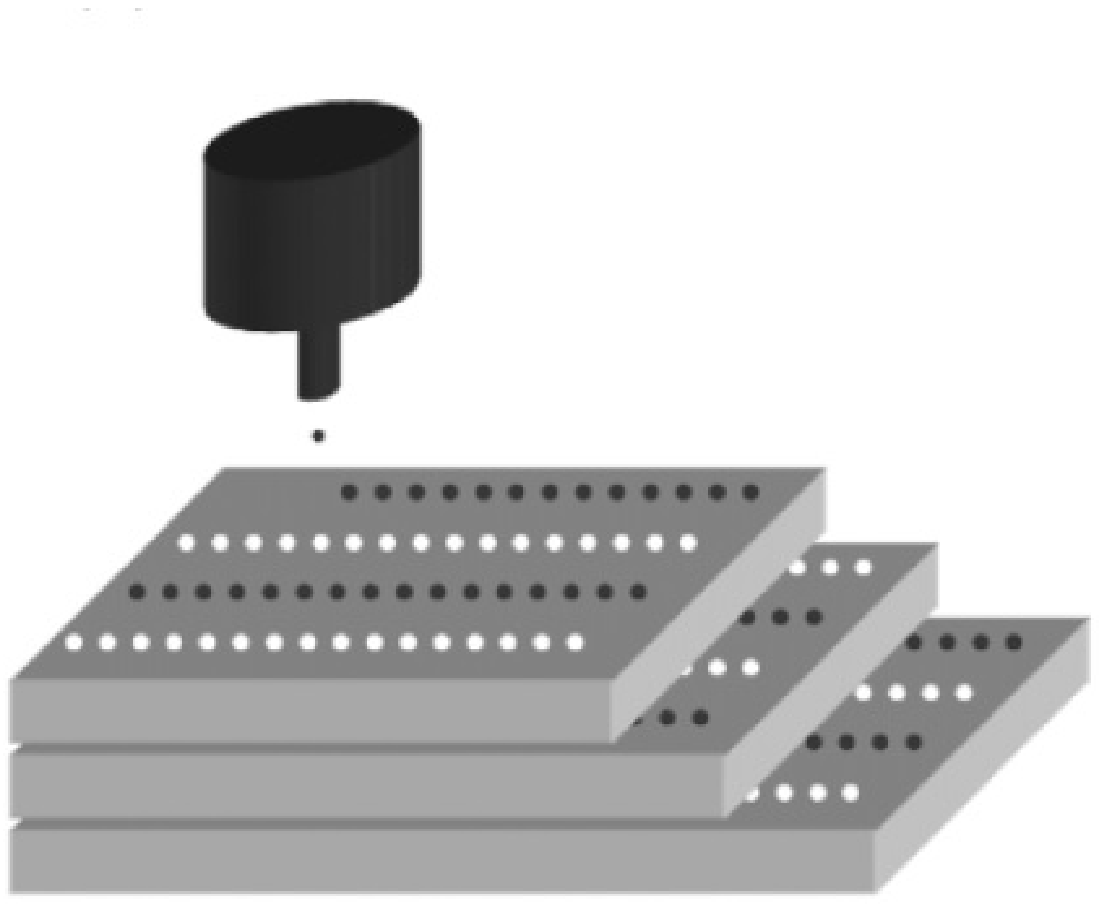}
                 \caption{}
                 \label{fig:scaffold2}
        \end{subfigure}
        ~ 
        \caption{Schematic representation of (a) scaffold-base printing when bio-materials and cells will be printed at the same time (b) cell printing where cells will be place on prefabricated scaffolds \cite{Ringeisen2006}.}\label{fig:scaffold}
\end{figure*}
Printing process for bio-printers can be categorized as two main groups: scaffold based printing and scaffold-free printing. As the names suggests, in scaffold-based printing, a skeleton of the organ/tissue geometry is first printed using certain bio-materials. The scaffold is then populated with living cells which will colonize the structure. At the end of the colonization process, the scaffold resolves into the systems.  Scaffold material porosity is designed in a way to enable inward diffusion of nutrients, oxygen and outward diffusion of waste materials from the living cells. Scaffold-based methods can further be categorized into two main approaches: first approach is when the scaffold will be printed out at the same time as the living cells (Figure \ref{fig:scaffold1}), and the other approach is when cells will be printed on layers of prefabricated scaffolds (Figure \ref{fig:scaffold2}) \cite{Ringeisen2006}.

In scaffold-free printing, living cells are directly printed onto a substrate. Natural cell processes such as cell sorting and cell fusion which occur through cell signaling automatically generate structure\cite{Beysens2000, Armstrong1989}. The manufacture of functional vasculature is a big challenge in bio-printing processes. Lack of vascular systems leads to cell death in systems where diffusion alone cannot handle transport of nutrients to cells and waste matter away from cells \cite{Luo2014}.

Both scaffold-base or scaffold-free methods use bio-printers that can be categorized into two general classes: Laser assisted bio-printing (LaBP) and Laser free bio-printing (LfBP). Each category contains multiple techniques that have been developed during the last couple of decades. In this review we will briefly address available printing methods and at the end a comparison between  different methods will be presented.

\section{Laser assisted bio-printers (LaBP)}

Laser assisted bio-printers (LaBP) were  the first type of bio-printers introduced by Odde and Renn \cite{Odde2000a} under name of "Laser-Guided Direct Writing" (LG DB). The printing setup included a laser beam, a focusing system and a substrate. The laser beam with the focusing system is used to create a "light trap" which is used to guide livig cells on the substracte to desired locations. The long and direct laser light contact with cells caused low cell survival rates\cite{Odde2000a}.  

\begin{figure*}[!t]
\centering
\includegraphics[width=0.8\linewidth]{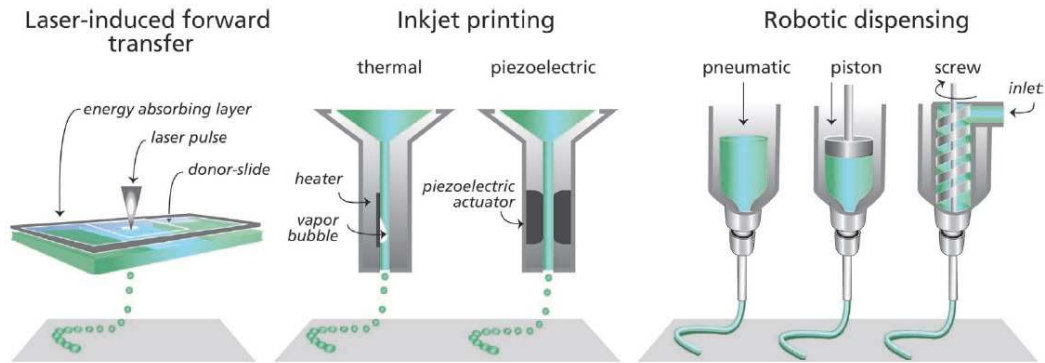}
\caption{Bio-printers with different printing approach. \cite{Malda2013}}
\label{fig:bioprinters}
\end{figure*}

Subsequently, the principle of laser-induced forward transfer (LIFT), initially developed for metals, has been adpated to bio-printing and is currently the main method for  LaBP \cite{Serra2006}. In a typical LaBP, which is printing based on LIFT method, (figure \ref{fig:bioprinters} ), focused laser pulses cause local vaporization of an energy absorbing layer (typically gold or titanium). This causes generation of a droplet based on the energy of the pulse and the duration of the pulse. The size of the droplet can be adjusted to meet the requirements of the process \cite{Guillemot2011,koch2013}. The generated droplets will be received by a substrate facing to the bio-ink \cite{Murphy2014,Malda2013}. With this configuration, the survival rate of cells was improved  drastically (up to 95\% cell viability) since the laser doesn't have direct contact with live cells \cite{Serra2006}. Simultaneously, the spatial resolution could be maintained in the range of $30-100\mu m$. The lack of a fluid orifice and reservoir in these designs means that issues such as print head clogging do not exist. However, the low printing speed (as low as $10^2$ drops/s) makes this process unsuitable for organ printing where the amount of material that must be deposited makes the turn around time prohibitive.

. 

\section{Laser free bio-printers (LfBP)}

\begin{figure}[!t]
\centering
\includegraphics[width=0.7\linewidth]{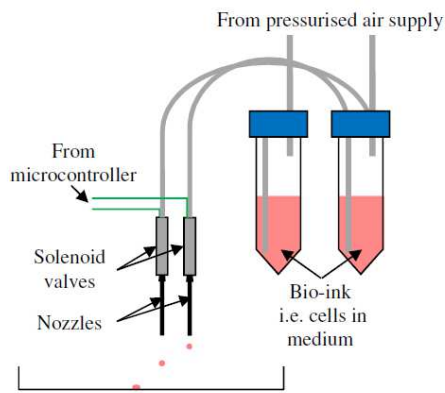}
\caption{Dispensing system for printer heads equipped with solenoid valves \cite{Faulkner-Jones2013}}
\label{fig:Solenoidvalves}
\end{figure}

Laser free bio-printers were inspired by inkjet desktop printers. Inkjet printers typically consists of a reservoir tank, micrometer size orifice, and a print head that can be actuated  either thermally, piezoelectrically (Figure \ref{fig:bioprinters}) or by solenoid valves (Figure \ref{fig:Solenoidvalves})  \cite{Faulkner-Jones2013, Malda2013} .  In inkjet printers, droplets are generated only when required. A pressure pulse is generated in the tank which force the bio-ink to go through the orifice which leads to the printer head. For printers that are equipped with a thermal print head, a micro-heater element vaporizes small pockets of the bio-ink to  produce vapor bubbles. Generation and collapse of these bubbles cause discontinuity in the bio-ink stream. In piezoelectric models, the print head is equipped with a piezoelectric actuator which deforms with a controlled frequency to make discontinuity in the fluid stream  \cite{Derby2008}. In the solenoid valve print head discontinuity in bio-fluid steam is be caused by a valve that can be turned on and off \cite{Faulkner-Jones2013}. In the all mentioned configurations, the print head is connected to a robotic base controlled by software. Therefore the print head can be accurately positioned over a substrate and the bio-ink droplet can be released at the desired location. 

In contrast with inkjet printers (generating droplets on demand) \cite{binder2011}, in robotic dispensing, there is no droplet generation process. Instead, the print head will produce a continuous stream of the bio-fluid and place it on the substrate. A controlled pressure system will forces the bio-ink to go through the nozzle. Depending on the viscosity of the fluid which requires different levels of the control, different pressure control systems can be utilized (Figure \ref{fig:bioprinters}).

The robotic dispensing method has the highest printing speed ($10-50 \mu m/s$) as well as high cell density. Among the LfBP systems, the robotic dispensing method has the lowest cell viability rates.($40-80\%$). Other LfBP are reported higher viability ($> 85 \%$) \cite{Ozbolat2013,Murphy2014,Malda2013}.

%

\section{Discussion }
There are several different configurations available for bio-printers. A end user has to choose and appropriate technique based on his or her requirements. Table \ref{table:comparison} illustrates a comparison between different bio-printing techniques that are currently available.

\begin{table*}[!t]
\centering
\begin{tabular}{lccccc}
\hline
Printer type & Resolution ($\mu m$) & Print speed & Cell viability(\%) & Cell densities &  References \\
\hline
LaBP &&&&\\
LG DW& 10-30& Continuous  ($ 9 \times 10^{-8} mL/s$) & Not Available as percentage & $~10^8 cells/ml$&  \cite{Odde2000a,Mironov2011,Jones2012}\\
Modified LIFT&30-100&  $10^2$ drops/s& 95-100& $~10^8 cells/ml$ & \cite{Serra2006,koch2013,Mironov2011,Jones2012}\\
\hline
LfBP &&&&\\
Thermal & $> 300$ &  $5 \times 10^3$ drops/s& 75-90& $<10^6 cells/ml$&  \cite{Derby2008,Mironov2011,Jones2012}\\
Piezoelectric &  - & $ 1 \times 10^4$ drops/s & ($> 85\% $) & $<10^6 cells/ml$&  \cite{Derby2008,Mironov2011,Jones2012}\\
Solenoid valve& - & 6500 drops/s & 85-99 & $~5 \times 10^5 cells/ml$&  \cite{Faulkner-Jones2013,Mironov2011,Jones2012}\\
Robotic dispensing& $ 5 \mu m$ to millimeters & $10-50 \mu m/s$& 40-80 \% & cell spheroids&  \cite{Xu2005,Guillotin2010,Mironov2011,Jones2012}\\ 
\hline
\end{tabular}
\caption{A brief comparison between bio-printers.}
\label{table:comparison}
\end{table*}

Laser guided direct write (LG DW) printer have the best spatial resolution and the lowest viability. LIFT design was a big step toward improvement of cells viability for laser assisted printers. Modification of LIFT (as presented in the table \ref{table:comparison} as modified LIFT) added an energy observing layer to the print head. This change had a huge positive impact on cell viability. Laser free printing methods sacrifice cell density and spatial resolution to achieve higher throughput. Print jobs that have intricate features are therefore more suitable for LaBP class of bio-printers. Print jobs with large tissue masses are more suited for LfBP methods.

\ifCLASSOPTIONcompsoc
  \section*{Acknowledgments}
\else
  \section*{Acknowledgment}
\fi

This work was partially supported by National Institutes of Health, NIH/IU Subcontract 2R01GM077138-04A1.

\ifCLASSOPTIONcaptionsoff
  \newpage
\fi



\bibliographystyle{IEEEtran}

\bibliography{IEEEabrv,bibio}

\end{document}